\documentclass[singlecolumn,notitlepage,pre,floatfix,longbibliography]{revtex4-1}
\usepackage{graphicx}
\usepackage[english]{babel}
\usepackage{amssymb}
\usepackage{amsbsy}
\usepackage{amsmath}
\usepackage{bbm}
\usepackage{xcolor}
\usepackage{hyperref}

\newcommand{\vx}{{\mathbf x}}

\newcommand{\vu}{{\mathbf u}}

\usepackage{natbib}
\begin{document}
\title{Solute mixing, dynamic uncertainty and effective dispersion in two-dimensional heterogeneous porous media}
 \author{Aronne Dell'Oca and Marco Dentz}
 \email[E-mail: ]{marco.dentz@csic.es}
 \affiliation{Spanish National Research Council (IDAEA-CSIC), Barcelona, Spain}
 
\date{\today}

\begin{abstract}
We study the mixing dynamics of a solute that is transported by advection and dispersion in a heterogeneous Darcy scale porous medium. We quantify mixing and dynamic uncertainty in terms of the mean squared solute concentration and the concentration variance. The latter measures the degree of mixing of the solute and at the same time the uncertainty around the mean concentration. Its evolution is controlled by the creation of concentration fluctuations due to solute spreading and its destruction by local dispersion. For moderate heterogeneity, this interplay can be quantified by using apparent and effective dispersion coefficients. For increasing heterogeneity we find deviations from the predicted behavior. In order to shed light on these behaviors and separate solute mixing and spreading, we decompose the solute plume into partial plumes, transport Green functions, and analyze their dynamics relative to those of the whole plume. This reveals that the variability in the dispersive scales of the Green functions in the plume and their interactions due to the strong focusing of preferential flow paths play an important role in highly heterogeneous porous media.  
\end{abstract}

\maketitle


\section{Introduction}
\label{section:Introduction}
Mixing is the process that drives the dilution of a solute by increasing its occupied volume, or area in two-dimensional settings, entailing the attenuation of the solute concentration content of a mixture (e.g., \citealt{dentz2022mixing}).  In addition, mixing is particularly important to control chemical reactions (e.g., \citealt{Valocchi2019, Rolle2019, dentz2011mixing}) that might occur when waters at different chemical equilibria come into contact, e.g., salty and fresh waters in coastal aquifers (\citealt{RivaSWI, Pool2015, DeVriendt2020}), riverine water and groundwater in the hyporheic zone (\citealt{NogueiraMixing, HypoMixWRR}). The proper quantification of solute dilution and reactivity is of primary interest in the context of contaminant risk assessments in subsurface systems (e.g., \citealt{Bellin2011, Fiori2001, Tartakovsky2013, dentz2010probability}).

\begin{figure}[t]
\begin{center}
\includegraphics[width=.55\textwidth]{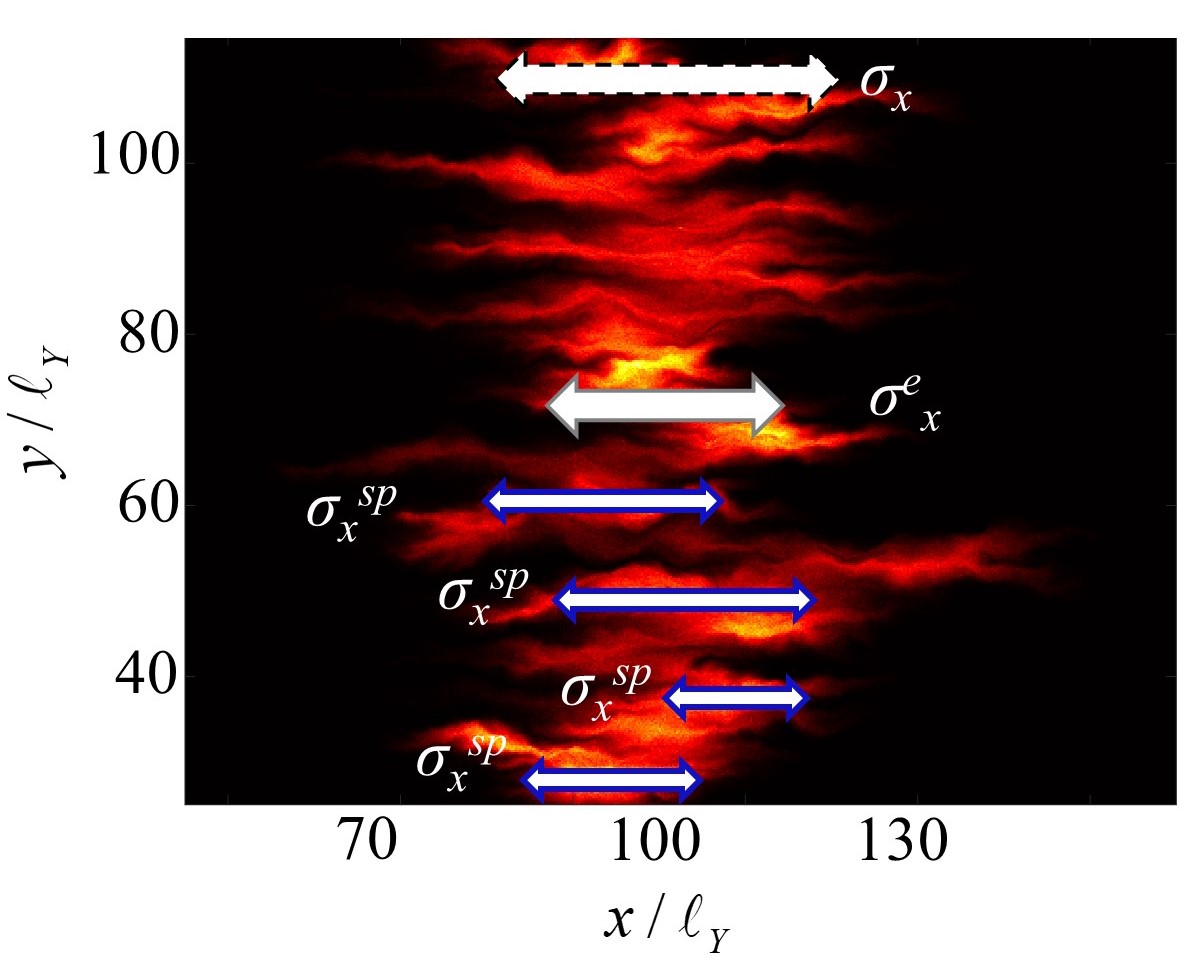}
\end{center}
\caption{Solute concentration field transported through a complex two-dimensional flow field in a porous medium over a hundred of heterogeneity characteristic length $\ell_Y$. The apparent dispersive scale $\sigma_x$ (double arrow with dashed border), the average dispersive scale $\sigma^{e}_x$ (double arrow with grey border) and a set of sub-plume dispersive scales $\sigma^{sp}_x$ (double arrows with blue borders) are sketched.} 
\label{fig:Intro}
\end{figure}

Mixing is ultimately carried out at the 'local' scale by diffusive mass transfers that are proportional to small-scale concentration gradients. In heterogeneous porous media,  the dynamics of the latter are controlled by the stirring and folding action of the flow field which enhances the dilution of the mixture (e.g., \citealt{deAnna2013, leborgneLamellar2015, BolsterHypermixing, dentz2022mixing}). At the same time, the heterogeneity-driven deforming action of the flow field sustains the overall dispersion of the solute mixture (e.g., \citealt{leborgneLamellar2015, Dagan84, Attinger, Dentz2000}). Figure \ref{fig:Intro} depicts the concentration distribution of a plume that has (on average) traveled several characteristic heterogeneity scales, i.e., $\ell_Y$, along the main flow direction $x$. The apparent dispersive scale $\sigma^a$ (double arrow with dashed border) imbues information about the large-scale dispersion of the mixture which only partially resolves mixing, e.g., \citealt{nonFickianMixing, BOLSTERstratified} 

Thus, the knowledge of $\sigma^a$ is insufficient to fully quantify mixing during time scales of practical interest when the mixture is not internally well-mixed. Downscaling the analysis of dispersion within the plume, it is possible to identify a distribution of sub-plume dispersive scales $\sigma_x^{sp}$ (double arrows with blue borders) and define their average $s^{e}$ (double arrow with grey border). The latter is the so-called effective dispersive scale (e.g., \citealt{Kitanidis88, DentzCarrera}) that is unbiased by the dispersion of the diverse sub-plume parcels that constitute the plume and evolves under the local interplay between the advective stirring action and the diffusive mass transfer. The connection between $\sigma_x^{e}$ and the mixing of a plume has been exploited to upscale the mixing dynamics at the pore and Darcy's (e.g., \citealt{cirpka2002choice, cirpkanowak, herrera2017lagrangian}) scales. Considering a small plume and resting on perturbation theory the statistical properties of Darcy's scale heterogeneous media have been linked to the time behavior of $s^{e}$ by \citealt{Dentz2000} and the mixing of a small blob by \citealt{DentzBlob}, while \citealt{Fiori2001} and \citealt{deBarrosFioriCDF} provide expression for the first two statistical moments and the whole cumulative distribution function of the peak concentration, respectively. Similarly,  \citealt{Vanderborght} leveraged on perturbation theory to predict the dilution of a large plume. These works provide valuable insights into the mixing dynamics of solute in Darcy's flow, but are restricted to moderate degrees of heterogeneity.

Considering mildly to highly heterogeneous geological formations,  \citealt{leborgneLamellar2015} adopt a lamellar description of mixing to capture the stretching of material elements, due to flow fluctuations, which control the dynamics of local concentration gradients and thus the intensity of diffusive mass transfers. Within this picture, the lamellae dilution by (transverse) diffusive expansion is simply superimposed on the stretched support, i.e., the diffusive sampling of the flow heterogeneity by the lamellae is disregarded. The latter makes the stretching histories of distinct lamellae independent, justifying the adoption of a random aggregation model to explain the late-time coalescence of lamellae as their support grows diffusely in the transverse direction. Recently, focusing on the pore scale, \citealt{Perez2023} modified the stretched lamellae approach to account for the impact of the diffusive sampling of the flow heterogeneity on a representative dispersive lamella. This approach has not been tested for transport in heterogeneous Darcy-scale media, and the possible implications of dispersive lamellae coalescence have not yet been addressed. 

Other conceptualizations and modeling strategies have been adopted to predict mixing in heterogeneous porous media. A trajectory-based mixing model has been employed in pore (\citealt{Sund2017}) and Darcy's (\citealt{wright2018upscaling}) scale analysis. This mixing model exploits the space-markovian nature of solute particle speeds over a certain characteristic scale $\ell_v$. This, jointly with the distribution of particle speeds, control the large-scale dispersion of the plume. The previously discussed non-mixed condition internally to the plume is captured by downscaling the longitudinal and transverse particle trajectories below $\ell_v$. In particular, the downscaling of the transverse trajectories requires the knowledge of a set of conditional probabilities. Alternatively, extensions of the interaction by exchange with the mean (IEM) mixing model (e.g., \citealt{pope_2000}) have been proposed on the common ground that mixing by porous media induces a multi-scale (e.g., \citealt{SoleMari2020}) and transient (e.g., \citealt{SCHULER, deDreuzy2012, kapoor1998}) mixing rate. The latter controls the relaxation properties of the concentration fluctuations (i.e., not well-mixed mixture content) towards the surrounding average concentration value. 

Overall, this set of studies on mixing in heterogeneous media suggests that we must recognize the key mechanisms that make the dilution of a mixture a sub-plume scale process. In this context, we highlight the relevance of retaining the sub-plume variability in the dispersive scales $\sigma^{sp}_x$, as well as, the coalescence dynamics between initially close (i.e., with correlated time evolutions) dispersive lamellae.

\section{Methodology}

\label{section:Flow and Transport Problems}
Heterogeneity is ubiquitous in environmental porous media spanning a vast fan of spatial scales that range from intricate pore architectures (e.g., \citealt{SienaGRL, AlecPRL, LayeredPORE}) to erratic variations in the hydraulic conductivity at the Darcy's (e.g., \citealt{sillimansimpson, DellOcaITPerm}) and regional scales (e.g., \citealt{NeumanDiFederico,  DellOcaITFlow}). Due to its relevance for practical applications at the field scale, we focus on heterogeneous hydraulic conductivity distribution at the Darcy's scale. In this section, we provide details about the the Darcy's scale flow and transport problems. 

\subsection{Flow and Transport Formulations}
\subsubsection{Flow model}
\label{subsection:Flow Problem}
We consider steady state Darcy's flow in a two-dimensional geological formation characterized by a heterogeneous distribution of the isotropic hydraulic conductivity tensor $\textbf{\textit{K}}(\vx)=K(\vx)\textbf{I}$, with $\vx=(x,y)$ being the spatial vector coordinates and $\textbf{I}$ the identity matrix. The effective porosity of the formation is considered homogeneous $\phi$. The Eulerian flow field $\textbf{\textit{v}}(\vx)$ is obtained by combining the Darcy's equation and the mass conservation principle, i.e., 

\begin{align}
\label{eq:flow}
\mathbf{{q}}(\mathbf{{x}}) = -{K}(\mathbf{{x}}) \nabla h (\mathbf{{x}}), 
\end{align}

where $h(\vx)$ is the hydraulic head. To rend the erratic nature of the hydraulic conductivity encountered in geological formations we treat $Y(\vx)=ln(K\vx/{K_g})$ as a second-order stationary multi-Gaussian random field characterized by the geometric mean $K_g$ and an isotropic exponential spatial covariance
\begin{align}
\label{eq:Ycovariance}
C_Y(\vx)=\sigma^2_Y e^{(-|\vx|/ l_Y)} 
\end{align}
where $\sigma^2_Y$ and $l_Y$ is the variance and the correlation length of $Y$, respectively. Note that, we identify $l_Y$ as the characteristic length of the problem and that $\langle \cdot \rangle$ will stand for the ensemble average with respect to the stochasticity in $Y$. Regarding the physical domain of interest $\omega$, we consider a rectangular region of dimensions $L=600l_Y$ and $H=150l_Y$ in the $x$ and $y$ directions, respectively. The hydraulic conductivity field is generated following a Sequential Gaussian Simulator scheme employing a regular Cartesian grid with element size $\Delta x=\Delta y=l_Y/10$. The investigated degree of formation heterogeneity spans from mildly to highly heterogeneous, i.e., $\sigma^2_Y=(1,4)$. To complement the flow problem \eqref{eq:flow} we impose a permeameter-like set of boundary conditions, i.e., no-flow along the bottom ($y=0$) and top ($y=150l_y$) boundaries, fixed hydraulic head $h(x=0,y)=h_{BC}$ and given velocity components $v(x=L,y)=(v_{x;BC},0)$ along the right and left boundary, respectively. Thus, we identify $x$ as the longitudinal (or mean flow) direction and $y$ as the transverse direction. Note that, at sufficient distances from the domain boundaries the ensemble average flow field is homogeneous and we identify the velocity module $\langle |\textbf{\textit{v}}| \rangle$ as the characteristic velocity of our problem. The flow field is obtained numerically through a mixed-finite element solver employing the same spatial grid associated with the generation of the hydraulic conductivity field. 

\subsubsection{Transport Model}
\label{subsection:Transport Problem}
We consider the transport problem at the Darcy's scale and governed by the advection-diffusion equation for the passive solute concentration $c(\vx,t)$
\begin{align}
\label{eq:DarcyTrans}
\frac{\partial c(\vx,t)}{\partial t}+\textbf{\textit{v}}(\vx)\cdot \nabla c(\vx,t)-D\nabla^2c(\vx,t)=0
\end{align}
where $t$ is time and $D$ is the local diffusive coefficient that is here assumed homogeneous and velocity independent for simplicity. As initial condition, we consider a straight solute plume transversally centered in the middle of the domain and placed at $30l_Y$ downstream from the left boundary. The initial plume extends over $L_y=90l_Y$ in the direction perpendicular to the main flow and has a uniform solute concentration, i.e., $c(\vx,t=0)=1/L_y$. 
The transport scenario is characterized by the Péclet number $Pe=\tau_D/\tau_a$ which is the ratio between the characteristic diffusive $\tau_D=l_Y^2/(2D)$ and advective $\tau_a=l_Y/\langle |\textbf{\textit{v}}| \rangle$ times considering with the characteristic length $l_Y$. 
The transport problem \ref{eq:DarcyTrans} is solved by relying on a random walk formulation in order to prevent numerical dispersion.  The solute concentration distribution is reconstructed through a dynamic binning scheme detailed in \ref{section: CBinning}. In the rest of the work, we refer to the results of the numerical solution of the flow and transport problems as direct numerical simulations (DNS).

\subsection{Solute Dispersion}
\label{section: Transport Observables}
In this Section, we describe the transport observables related to the dispersion of solutes that are of interest to our study. In general, the analysis of solute dispersion refers to the characterization of the spatial extension of a solute plume.

\subsubsection{Apparent Dispersive Scale}
\label{subsection:Appdispersive}
Considering the longitudinal direction, the dispersive scale of a solute plume $\sigma_x(t)$ can be identified as the square root of the second centered spatial moment of the plume concentration field, i.e.,
\begin{align}
\label{eq:SpreadScale}
\sigma_x(t)=\sqrt{x_{2}(t)-x_{1}(t)^2}
\end{align}
where $x_{1}(t)$ is the solute plume longitudinal centroid and $x_{2}(t)$ is the second spatial moment of the plume, i.e.,
\begin{align}
\label{eq:Plume_m1_m2}
x_{1}(t)= \int\limits_\Omega  d\vx x c(\vx,t), && x_{2}(t)= \int\limits_\Omega  d\vx x^2 c(\vx,t)
\end{align}

\subsubsection{Dispersion of Green functions centroids}
\label{subsection:LPdispersive}
The concentration distribution of a solute plume can be expressed in terms of the corresponding Green functions as (e.g., \citealt{DentzdeBarrosVariance2013} )
\begin{align}
\label{eq:CPlumeGreens}
c(\vx,t)=\int\limits_{L_y}dy'    g(\vx,t|y')
\end{align}
where the Green function $g(\vx,t|y')$ (GF) satisfies Equation \ref{eq:DarcyTrans} considering the point-like initial condition $g(\vx,t=0|y')=\delta(x-30\ell_Y)\delta(y-y')$. 
The dispersion of the GFs centroids in the longitudinal direction is quantified by  
\begin{align}
\label{eq:DispersionCMs}
\sigma_{x_1^g}(t)=\sqrt{x_{2}^{g;e}(t)-x_{1}^{g;e}(t)^2}
\end{align}
where $x_{1}^{g;e}(t)$ and  $x_{2}^{g;e}(t)$ read 
\begin{align}
\label{eq:averageCMs}
x_{1}^{g;e}(t)=\frac{1}{L_y} \int\limits_{L_y}  dy' x_1^g(t|y'), && 
x_{2}^{g;e}(t)= \frac{1}{L_y} \int\limits_{L_y}  dy' (x_1^{g}(t|y'))^2
\end{align}
For the sake of subsequent discussions, we introduce the transverse coordinate of a GF centroid 
\begin{align}
\label{eq:rGF_m1_y}
y_{1}^g(t|y')= \int\limits_\Omega  d\vx y g(\vx,t|y')
\end{align}


\subsubsection{Effective Dispersive Scale}
\label{subsection:Effdispersive}
The dispersive scale of a GF is defined as 
\begin{align}
\label{eq:SpreadScale_rGF}
\sigma_{x}^g(t|y')=\sqrt{x_{2}^g(t|y')-x_{1}^g(t|y')^2}
\end{align}
where the centroid and second spatial moment of $g(\vx,t|y')$ are defined as
\begin{align}
\label{eq:rGF_m1_m2}
x_{1}^g(t|y')= \int\limits_\Omega  d\vx x g(\vx,t|y'), && 
x_{2}^g(t|y')= \int\limits_\Omega  d\vx x^2 g(\vx,t|y')
\end{align}
The effective dispersive scale $\sigma_{x}^e(t)$ of a solute plume is defined by averaging Equation \ref{eq:SpreadScale_rGF} over the initial condition, i.e., 
\begin{align}
\label{eq:Eff_SpreadScale}
\sigma_{x}^e(t)=\frac{1}{L_y} \int\limits_{L_y}dy'   \sigma_{x}^g(t|y')  
\end{align}

\subsection{Solute Mixing and Dynamic Uncertainty}
\subsubsection{Mixing State}

We define the mixing state of the plume as the integral over the domain of the concentration squared (\citealt{nonFickianMixing, BOLSTERstratified})
\begin{align}
\label{eq:M}
    M(t)= \int\limits_\Omega  d\vx c^2(\vx,t); 
\end{align}
We derive in APPENDIX B the following evolution equation for $M(t)$ (see also \citealt{kapoor1998})
\begin{align}
\label{eq:ScalarDiss}
    \frac{d M(t)}{dt} = - 2 \chi(t);   && \chi(t) =\int d\vx  D[\nabla c(\vx,t)]^2
\end{align}
where $M(t)$ is the mixing state and $\chi(t)$ is the scalar dissipation rate. In our work, we focus on ${M}(t)$ as a global descriptor of the mixing process. 

For a line source in a homogeneous two-dimensional medium, the concentration distribution is given by 
\begin{align}
    c(\vx,t) = \frac{1}{L_y}\frac{\exp[- (x- u t)^2/4Dt]}{\sqrt{4\pi D t}}. 
\end{align}
Using this expression in Eq.~\eqref{eq:ScalarDiss} gives
\begin{align}
   M(t) = \frac{t^{-1/2}}{4 L_y \sqrt{2  \pi D}} ; &&  \chi(t) = \frac{t^{-3/2}}{8 L_y \sqrt{2 \pi D}}. 
\end{align}

\subsubsection{Concentration Average and Variance}
\label{subsection:CAverage}
We consider the spatial average of the concentration field along the transverse direction 
\begin{align}
\label{eq:CSpatialAverage}
\overline{c}(x,t)=\frac{1}{L_y} \int\limits_{L_y} dy c(\vx,t)
\end{align}
and the corresponding concentration fluctuations field 
\begin{align}
\label{eq:C Ave Fluct}
c'(\vx,t)=c(\vx,t) - \overline{c}(x,t)
\end{align}
The concentration variance is defined as
\begin{align}
\label{eq:varC}
\sigma^2_c (x,t)=\frac{1}{L_y} \int\limits_{L_y} dy c'(\vx,t)^2
\end{align}


The concentration variance quantifies the uncertainty in the concentration values along the transverse direction around the average concentration. We define the overall uncertainty in the concentration value by the spatial integral of the concentration variance 
\begin{align}
\label{eq:Mu}
M_u(t)= L_y\int dx  \sigma^2_c(x,t) 
\end{align}
This is related to the mixing state as 
\begin{align}
\label{eq:M}
M(t)=M_a(t)+M_u(t); &&
M_a(t)=L_y\int dx \overline{c}^2(x,t);
\end{align}
where $M_a(t)$ quantifies the amount of mixing ascribable to the apparent spreading of the plume $\sigma_x(t)$. The contribution $M_u(t)$ reflects the presence of internal concentration fluctuations $c'(\vx,t)$ which dictate the deviation from well-mixed conditions at the plume scale. The concentration uncertainty evolves in time just like the mixing state according to (  \citealt{kapoor1998})
\begin{align}
 \frac{d M_u(t)}{dt} = - 2 \chi'(t);    && \chi'(t) =\int dx D\overline{  \frac{\partial c'(\vx,t)}{\partial x}^2 }   
\end{align}
A largely adopted closure approximation, derived in the context of turbulent flows, is the interaction by exchange with the mean (IEM) \citealt{pope_2000}, i.e., 
\begin{align}
 \overline{  \frac{\partial c'(\vx,t)}{\partial x}^2 }   \approx   \frac{\sigma^2_c (x,t) }{\lambda_c^2}   
\end{align}
where $\lambda_c$ is called the concentration microscale and was originally proposed to be a constant. The latter results in an exponential decay of the dynamic uncertainty $M_u(t)$ in time. On the other hand, investigations of the mixing dynamics in Darcy-scale heterogeneous porous media suggest a more complex behavior due to the time dependence of $\lambda_c(t)$ under the action of fluid deformation and diffusion (\citealt{ kapoor1998, Andrevick, deDreuzy2012, SCHULER, LEBORGNE2010, SoleMari2020}).  

\subsection{Dispersive Lamella Mixing Model}
\label{subsection: Dispersive Lamellae}
We briefly recall the basis of the dispersive lamella mixing model of \citealt{Perez2023}.

which suggests that the mixing problem can be resolved by knowing the area occupied by the GFs of the plume. The dispersive lamellae approach introduces a representative Green function approximated by a Gaussian distribution that undergoes dispersion along the longitudinal direction according to the effective dispersive scale $\sigma_x^e(t)$, i.e., 
\begin{align}
\label{eq:DL}
g(\vx,t)= g^e_x(x,t)g^e_y(y,t|y'); && 
g^e_x(x,t)= \frac{e^{-\frac{1}{2} \left( \frac{x-x_{1}(t)}{\sigma^{e}_x(t)} \right)^2 }}{\sigma^{e}_x(t)\sqrt{2\pi}}; && g^e_y(y,t|y')=\frac{1}{L_y}\delta(y-y')
\end{align} 
Note that, $g^e_x(x,t)$ is centered at $_{1}^(t)$ for convenience and that $g^e_y(y,t|y')$ simplifies to a Dirac's delta since the transverse dispersion of is smaller than the longitudinal counterpart. At the same time, the dilution of $g(\vx,t)$ due to the transverse diffusive sampling of the velocity fluctuations is encompassed by $\sigma^{e}_x(t)$.
Combining \eqref{eq:DL} and \eqref{eq:CPlumeGreens} and recalling the definition of the mixing state in \eqref{eq:ScalarDiss} leads to
\begin{align}
\label{eq:MixingGreensEffective}
M^e(t)= \int\limits_\Omega d\vx g(\vx,t|y') ^{2}=\frac{1}{L_y}\frac{1}{\sqrt{\pi}2\sigma_x^e(t)}
\end{align}

\section{Results}
\label{subsection:Results}
In this section, we inspect the dispersive and mixing dynamics for a large plume that travels in mildly, Section \ref{subsection:Low Het}, and highly, Section \ref{subsection:High Het}, heterogeneous formations at Darcy's scale and we discuss the capability of the dispersive lamella mixing model, Section \ref{subsection: Dispersive Lamellae}, to capture the mixing dynamics in these scenarios. 

\subsection{Mildly Heterogeneous Formation}
\label{subsection:Low Het}

\begin{figure}[t]
\begin{center}
\includegraphics[width=.85\textwidth]{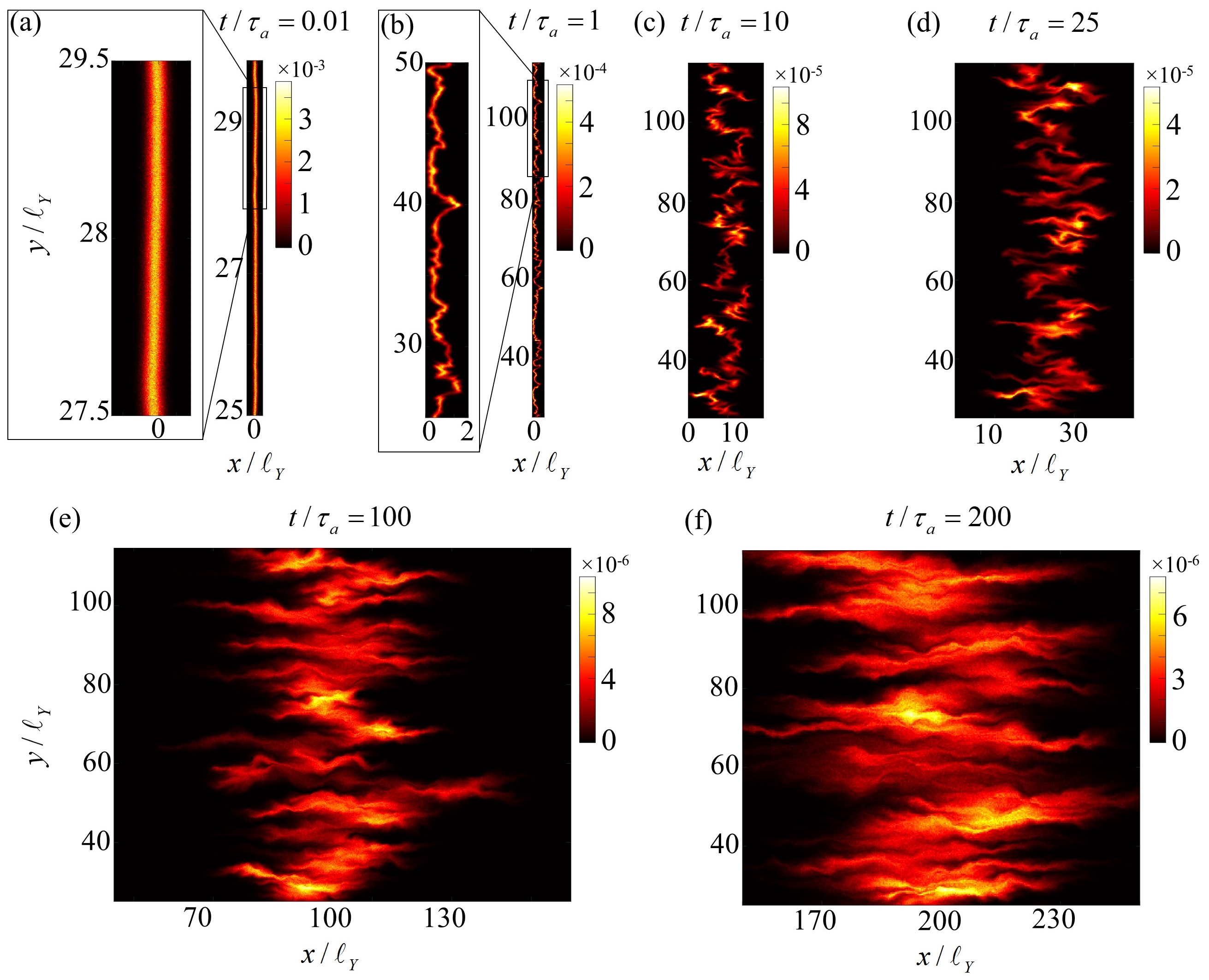}
\end{center}
\caption{Solute concentration field at diverse times $t/\tau_a$ considering a mildly heterogeneous distribution of the hydraulic conductivity field, $\sigma_Y^2=1$, and an advective dominated transport scenario, $Pe=100$.} 
\label{fig:Plume Times}
\end{figure}

Considering a mildly heterogeneous formation $\sigma^2_Y=1$, Figure \ref{fig:Plume Times} depicts the evolution of the  plume concentration field over times $t/\tau_a$: (a) at early times local diffusion dictates the longitudinal growth of the initially straight line; (b-d) as time passes, the variability in the advective component of transport distorts the plume, stretching the plume sub-parcels in the longitudinal direction which favors the sampling of the flow variability by transverse diffusion; (e-f) at late times, the plume tends to homogenize and to approach well-mixed conditions. 

\begin{figure}[t]
\begin{center}
\includegraphics[width=.55\textwidth]{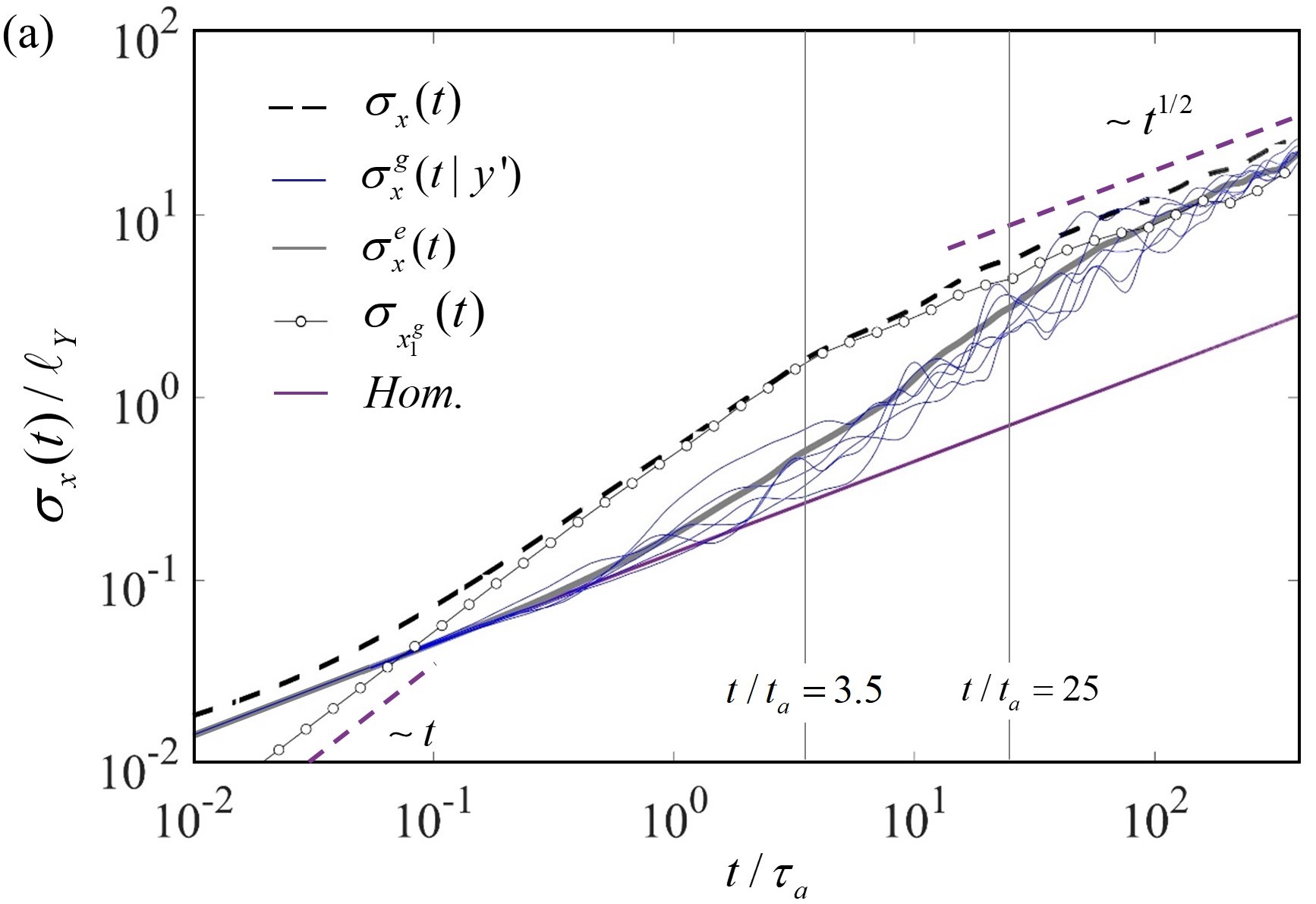}
\end{center}
\caption{Dispersive scales: $\sigma_x(t)$ (dashed black curve), $\sigma_x^e(t)$ (grey curve), $\sigma_x^g(t|y')$ (blue curves) for a set of Green function and $\sigma_{x_1^g}(t)$ (white circles) for $\sigma_Y^2=1$ and $Pe=100$. The homogeneous case (purple lines) and the Fickian scaling (dashed purple lines) are depicted as references.} 
\label{fig:Spreading Scales SigY = 1}
\end{figure}

Figure \ref{fig:Spreading Scales SigY = 1} depicts the apparent dispersive scale $\sigma_x(t)$ (dashed black curve), effective dispersive scale $\sigma_x^e(t)$ (grey curve), dispersive scales for a set of Green functions $\sigma_x^g(t|y')$ (blue curves) and the dispersive scale of the Green functions centroids along the main flow direction $\sigma_{x_1^g}(t)$ (white circles). Inspection of Figure \ref{fig:Spreading Scales SigY = 1} highlights that  $\sigma_x^e(t)$ grows according to $\sqrt{Dt}$ since the majority of the GFs are not yet large enough to experiment the variability in the flow field. The latter are subsequently, i.e., $t/\tau_a>1$, experimented by GFs due to local transverse diffusive mass transfers and $\sigma_x^e(t)$ follows a super-diffusive regime. At late times, i.e., $t/\tau_a>100$, the diverse GFs have grwon sufficently large to experiment the whole variability in the flow and thus $\sigma_x^e(t)$ approaches the large-scale Fickian diseprsvie regime, without reaching it in the considered time window (see also \citealt{Dentz2000}). At the same time, the apparent dispersive scale $\sigma_x(t)$ generally overestimates $\sigma_x^e(t)$, until late times when the two dispersive scales tend to coincide. The discrepancies between $\sigma_x(t)$ and $\sigma_x^e(t)$ are ascribable to the dispersion of GFs centroids $\sigma_{x_1^g}(t)$. Until $t/\tau_a<3.5$ (approximately), $\sigma_{x_1^g}(t)$ grows ballistically, i.e., the GFs are small objects whose dispersion of centroids resemble that of purely advected particles. Subsequently, for $t/\tau_a > 3.5$, $\sigma_{x_1^g}(t)$ (and $\sigma_{x}(t)$) grows at a slower pace: similar to purely advective transport, initially close GFs are brought closer to each other to form a sub-set of GFs that is majorly elongated in the longitudinal direction as it is conveyed towards the nearest preferential flow path. Note that, $t/\tau_a\approx3.5$ is the time at which the transverse dispersion coefficient of the purely advected particles peaks for $\sigma^2_Y=1$ (e.g., \citealt{Jankovic2009}), corroborating the similarity between the motion of the GFs centroids and purely advected particles over short travel distances. The transport of initially close GFs into the nearest preferential flow paths increases the degree of correlation in the dispersive behaviors of the sub-sets of GFs, since GFs in a given sub-set experience a very similar series of flow fluctuations downstream (diffusion will eventually destroy this correlation at late times). This, in conjunction with the homogenization of the speeds of the diverse GFs centroids by the diffusive sampling of the flow heterogeneity, favors the slowdown in the growth of $\sigma_{x_1^g}(t)$. The latter is particularly evident in Figure \ref{fig:Spreading Scales SigY = 1} for $t/\tau_a>25$ (approximately), after which $\sigma_x(t)$ approachies the large-scale Fickian regime, i.e., $\sigma_x(t) \sim t^{1/2}$. The dominance of advection on the pre-asymptotic times for the dispersion of GFs centroids is corroborated by (\textit{i}) \citealt{Comolli2019} that suggested the relaxation time for the ensemble longitudinal dispersion of purely advected particles close to $25\tau_a$, and (\textit{ii}) \citealt{drDreuzy} that reported a strong similarity in the apparent longitudinal dispersive dynamics between the purely advective case and the counterpart for high Péclet in case of $\sigma^2_Y=1$. The stabilization of the GFs dispersion, after the longitudinal relaxation time, is key to the approaching of the Fickian scaling in both $\sigma_x(t)$ and $\sigma_x^e(t)$.

\begin{figure}[t]
\begin{center}
\includegraphics[width=.55\textwidth]{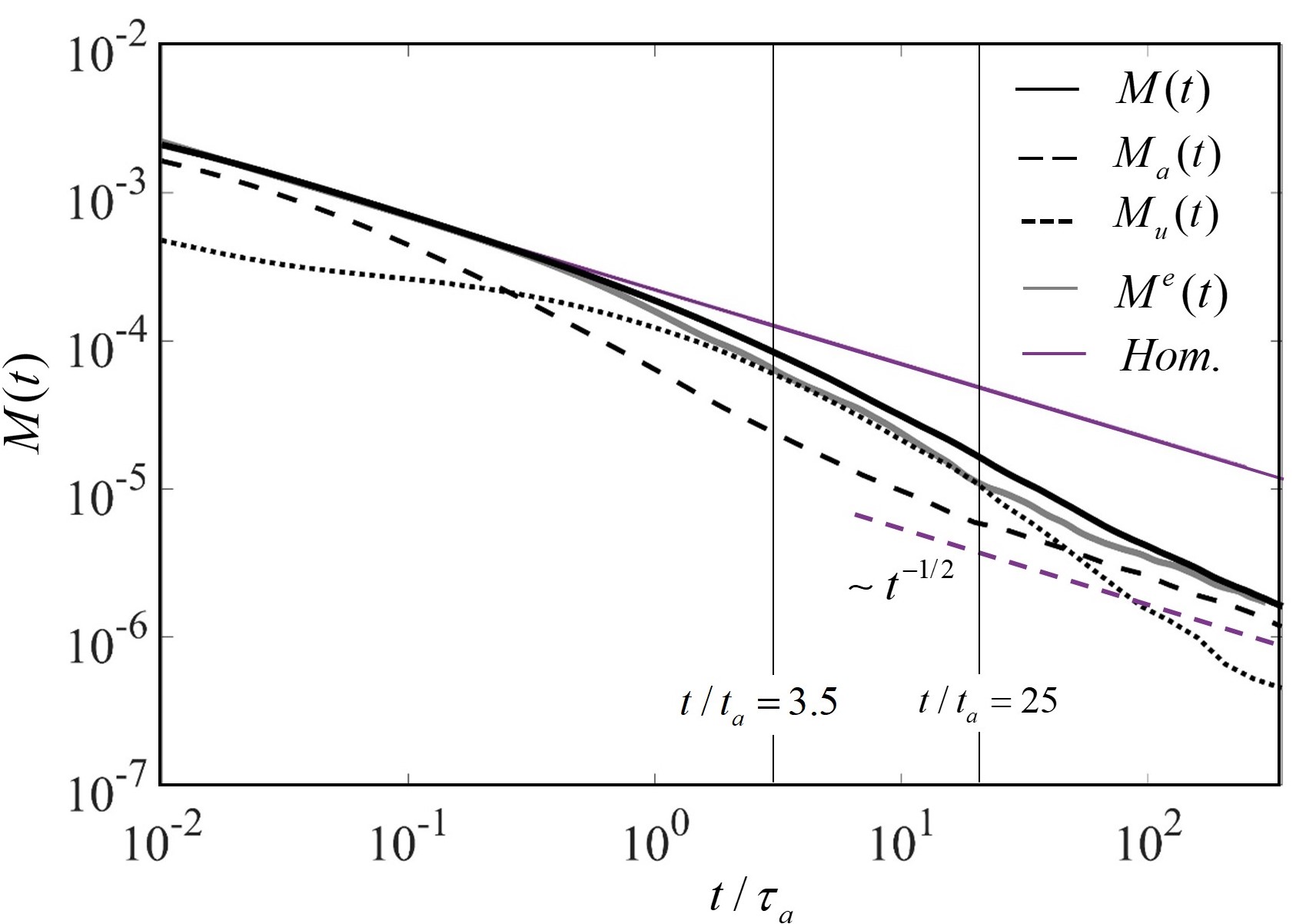}
\end{center}
\caption{Plume mixing state grounded on DNS results $M(t)$ (solid black curve)and its contributions $M_a(t)$ (dashed black curve) and $M_u(t)$ (dotted black curve). The mixing state grounded on the effective dispersive scale $M^e(t)$ (grey curve) is also depicted.  Results are for $\sigma_Y^2=1$ and $Pe=100$.  The homogeneous case (purple lines) and the Fickian scaling (dashed purple lines) are depicted as references.} 
\label{fig:Mixing State  SigY = 1}
\end{figure}

Figure \ref{fig:Mixing State  SigY = 1} depicts the mixing state as grounded on DNS result $M(t)$, highlighting the $M_a(t)$ (dashed black curve) and $M_u(t)$ (dotted black curve) contributions, and the prediction of the dispersive lamella mixing model $M^e(t)$ (grey curve). Inspection of Figure \ref{fig:Mixing State  SigY = 1} highlights that $M_a(t)$ generally overestimates the dilution of the solute plume, since it predicts that the plume is well-mixed over the apparent dispersive scales $\sigma_x(t)$.

At the same time, the dynamic uncertainty $M_u(t)$ dominates the mixing of the plume over the pre-asymptotic times during which the dispersion of the GFs centroids, see previous discussion of $\sigma_{x_1^g}(t)$, favors the internal segregarion of the plume and thus incomplete mixing at the plume scale. Note that, the discrepancy between $M_u(t)$ and $M_a(t)$ peaks around  $t/\tau_a<3.5$ (approximately), i.e., the time scale at which sub-sets  of initially close GFs are conveyed into the nearest preferential flow paths enhancing thus lacunarities in the plume. Additionally, $M_u(t)$ markedly decreses after the relaxation time $t/\tau_a=25$, i.e., when the motion of the GFs centroids has been sufficiently regularized (e.g., straight trajectories at more uniform speeds) to allow for local mass transfers to efficiently homogenize the plume internally. At the same time, $M_a(t)$ approaches the Fickian mixing scaling around $t/\tau_a=25$, i.e., when $\sigma_x(t)$ grows only under the amount of flow variability sampled by the GFs and not by the dispersion of their centroids. At late times, $M_a(t)$ approaches the plume mixing state $M(t)$ that progressively becomes well-mixed at the plume scale but is persistently non-Fickian due to the non-negligible contribution of $M_u(t)$ over the inspect time window, consistently with \citealt{nonFickianMixing},

Inspection of \ref{fig:Mixing State  SigY = 1} suggests that the dispersive lamella approach reflects the evolution of $\sigma_x^e(t)$, i.e., a purely diffusive scaling at early times followed by the advection-enhanced mixing regime during $1<t/\tau_a<25$. Moreover, we found that $M^e(t)$ captures the salient mixing mechanisms that drive $M(t)$ in mildly heterogeneous formations. Note that, around $t/\tau_a>1$ the  dispersive lamella overestimates the degree of mixing of the plume $M^e(t) < M(t)$. At the same time, the inspection of \ref{fig:Spreading Scales SigY = 1} reveals a certain degree of variability in $\sigma_x^g(t|y')$ around $t/\tau_a>1$, i.e., there is a sub-plume scale variability in the growing of the mixing area. Interestingly, after the relaxation time $ t/\tau_a>25$ the overestimation of the degree of mixing tends to reduce and $M^e(t)$ approaches $M(t)$ consistently the previous discussion about the evolution of the plume GFs.

\subsection{Highly Heterogeneous Media}
\label{subsection:High Het}
We proceed to analyze the dispersive and mixing dynamics in a highly heterogeneous formation, i.e., $\sigma^2_Y=4$.

\begin{figure}[t]
\begin{center}
\includegraphics[width=.55\textwidth]{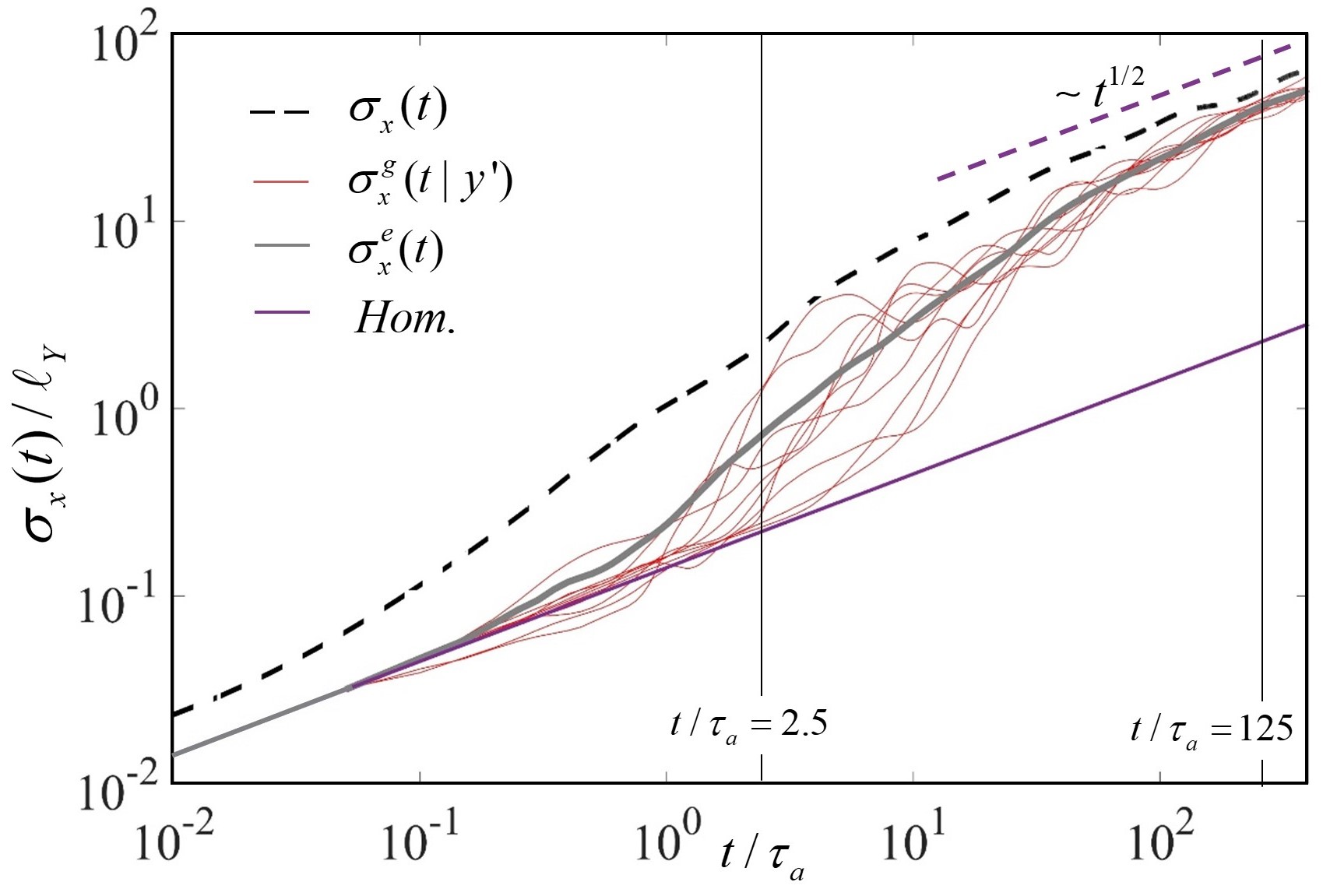}
\end{center}
\caption{Dispersive scales: $\sigma_x(t)$ (dashed black curve), $\sigma_x^e(t)$ (grey curve), $\sigma_x^g(t|y')$ (blue curves) for a set of Green function and $\sigma_{x_1^g}(t)$ (white circles) for $\sigma_Y^2=4$ and $Pe=100$. The homogeneous case (purple lines) and the Fickian scaling (dashed purple lines) are depicted as references.} 
\label{fig:Spreading Scales SigY = 4}
\end{figure}

Figure \ref{fig:Spreading Scales SigY = 4} depicts $\sigma_x(t)$ (black dashed curve), $\sigma_x^e(t)$ (grey curve) and a set of $\sigma_x^g(t|y')$ (red curves). Comparison of Figure \ref{fig:Spreading Scales SigY = 4} and Figure \ref{fig:Spreading Scales SigY = 1}a highlights that $\sigma_x(t)$ and $\sigma_x^e(t)$ are enhanced by heterogeneity, as well as, their relative discrepancies over time, as expected. Furthermore, we note a change in the slope of $\sigma_x(t)$ and $\sigma_x^e(t)$ around  $t/\tau_a=2.5$, i.e., the peak time of the transverse dispersion in case of pure advection (see  \citealt{DellOcaDentzINITIAL}), and $\sigma_x(t)$ approaching the Fickian scaling around $t/\tau_a=125$, i.e., the relaxation time of the longitudinal dispersion coefficient (see \citealt{Comolli2019, DellOcaDentzINITIAL}). Overall, this comparison hints at the similarity in the spreading dynamics for $\sigma^2_Y=1$ and $\sigma^2_Y=4$ suggesting the adoption of the dispersive lamellae mixing model to capture the mixing dynamics in highly heterogeneous formations.

\begin{figure}[t]
\begin{center}
\includegraphics[width=.55\textwidth]{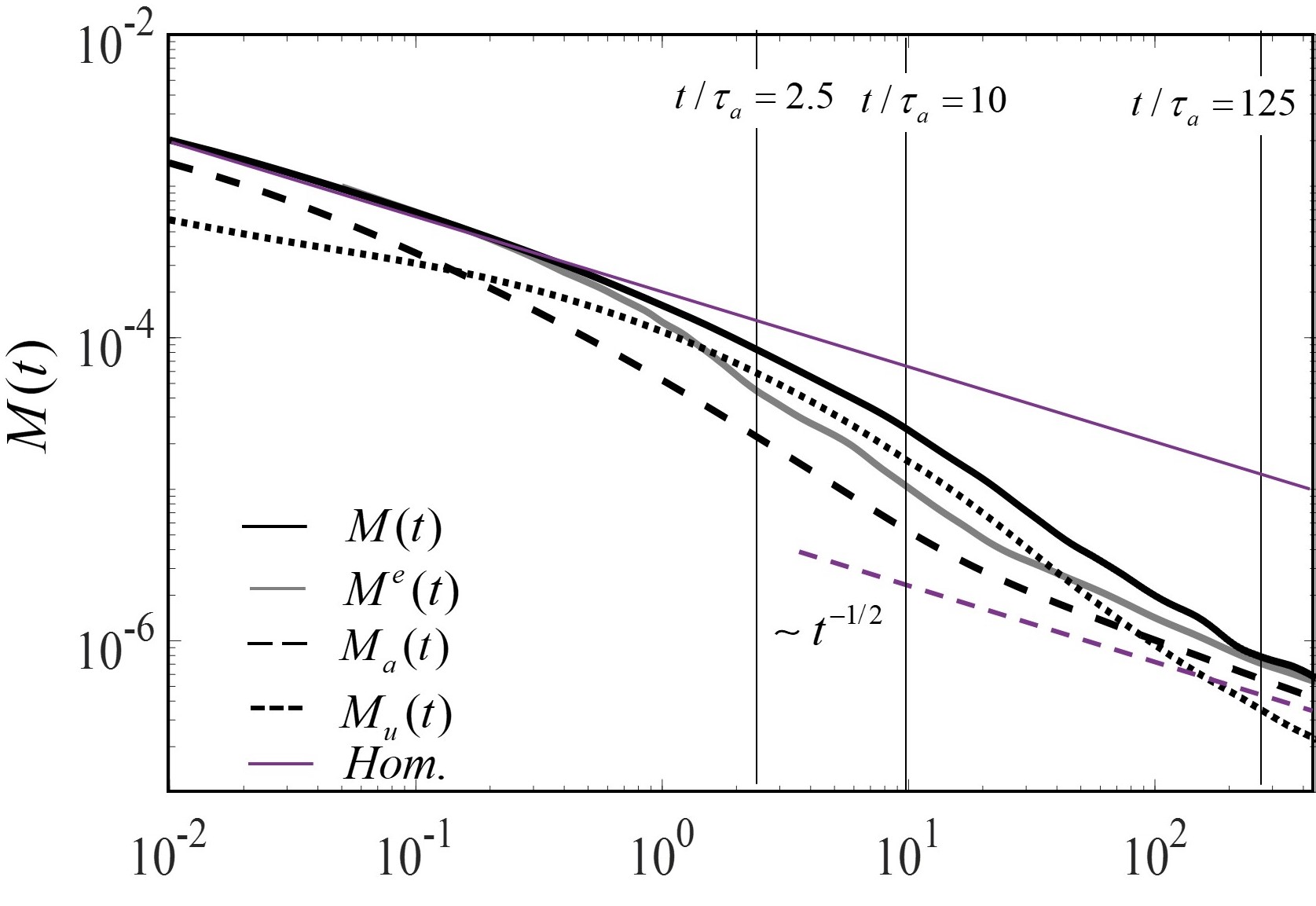}
\end{center}
\caption{Plume mixing state grounded on DNS results $M(t)$ (solid black curve)and its contributions $M_a(t)$ (dashed black curve) and $M_u(t)$ (dotted black curve). The mixing state grounded on the effective dispersive scale $M^e(t)$ (grey curve) is also depicted.  Results are for $\sigma_Y^2=4$ and $Pe=100$.  The homogeneous case (purple lines) and the Fickian scaling (dashed purple lines) are depicted as references.} 
\label{fig:Mixing State  SigY = 4}
\end{figure}

Figure \ref{fig:Mixing State  SigY = 4} juxtaposes $M(t)$ (black curve) grounded on the DNS and $M^e(t)$ (grey curve) for a highly heterogeneous formation: (\textit{i}) $M^e(t)$ clearly overestimates the degree of mixing as the flow fluctuations enhance the dilution of the plume; (\textit{ii}) $M^e(t)$ tends toward $M(t)$ at late times, $t/\tau_a>125$, in agreement with the discussion in \ref{subsection:Low Het}. Thus, despite the similarity in the dynamics of $\sigma_x^e(t)$ between $\sigma^2_Y=1$ and $\sigma^2_Y=4$ settings the upscaling of mixing grounded on the spreading lamellae has notable limitations over the pre-asymptotic times in case of highly heterogeneous formations.

\begin{figure}[t]
\begin{center}
\includegraphics[width=.55\textwidth]{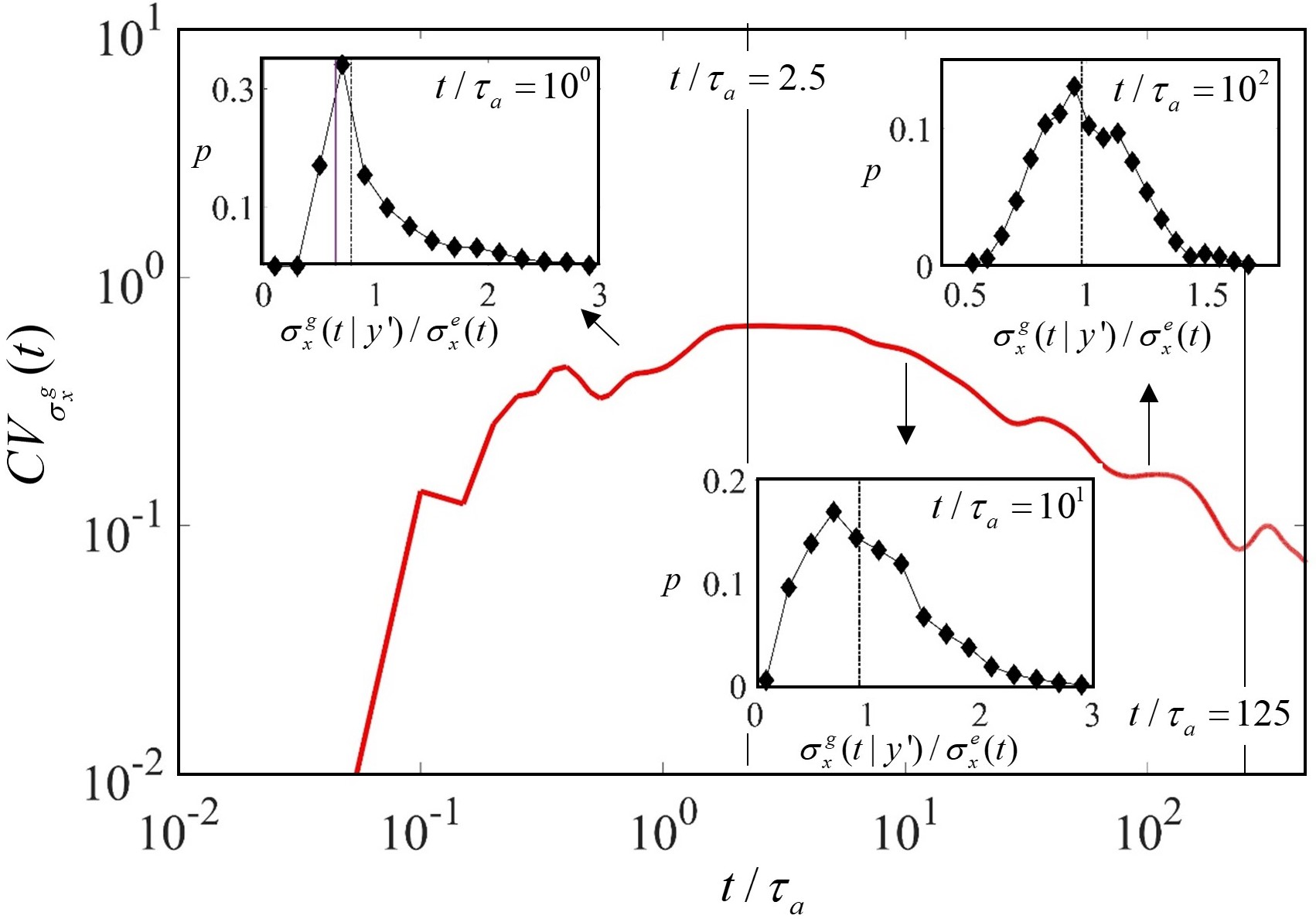}
\end{center}
\caption{The coefficient of variation $CV_{\sigma_x^g}(t)$ (red curve) and the distribution $p_{\sigma_x(t|y')}$ at given times (see insets) jointly with the median values (dashed black line). In the inset for $t/\tau_a=1$, the ratio $\sqrt{2Dt}/\sigma_x^e(t)$ (purple line) is also depicted. Results are for a highly heterogeneous formation,i.e., $\sigma_Y^2=4$ and $Pe=100$. } 
\label{fig:CV Spreading Scales SigY = 4}
\end{figure}

Inspection of Figure \ref{fig:Spreading Scales SigY = 4} highlights the variability of $\sigma_x(t|y')$ (red curves) around $\sigma_x^e(t)$ (grey curve) which is no longer sufficient to fully capture the mixing state over the pre-asymptotic times in case of a highly heterogeneous formation. We quantify the relative discrepancies of the sub-plume scale spreading scales $\sigma_x(t|y')$ in \ref{fig:CV Spreading Scales SigY = 4} which depicts (\textit{i}) the coefficient of variation $CV_{\sigma_x^g}(t)$ (grounded on the spatial statistics of the GFs constituting the plume) versus time and (\textit{ii}) the distribution $p_{\sigma_x(t|y')}$ (insets, the dashed vertical line is the median value of the distribution) at $t/\tau_a=(1,10,100)$. For $t/\tau_a=1$ we depict also the ratio $\sqrt{2Dt}/\sigma_x^e(t)$ (purple line). Considering $t/\tau_a=1$, we note that the median value of $p_{\sigma_x(t|y')}$ is smaller than $\sigma_x^e(t)$ and close to the purely diffusive value (purple line in the inset), i.e., at early times the majority of GFs are compact objects that mainly dilute under diffusion and not according to $\sigma_x^e(t)$. On the other hand, $p_{\sigma_x(t|y')}$ at $t/\tau_a=1$ exhibits a long tail for positive values which suggests an overall deviation from the purely diffusive mixing regime. Inspection of $CV_{\sigma_x^g}(t)$ reveals that an intense growth until $t/\tau_a=2.5$ followed by a plateau until $t/\tau_a\approx10$, the previously mentioned advection-dominated transport of initially close GFs towards the nearest preferential flow path imprint very similar evolutions in their dispersive scales $\sigma_x(t|y')$. After $t/\tau_a\approx10$, $CV_{\sigma_x^g}(t)$ declines due to the sampling of the flow heterogeneity by the diverse GFs in the plume. This is also evident from the less skewed behavior of $p_{\sigma_x(t|y')}$ for $t/\tau_a=(10; 100)$.

\begin{figure}[t]
\begin{center}
\includegraphics[width=.90\textwidth]{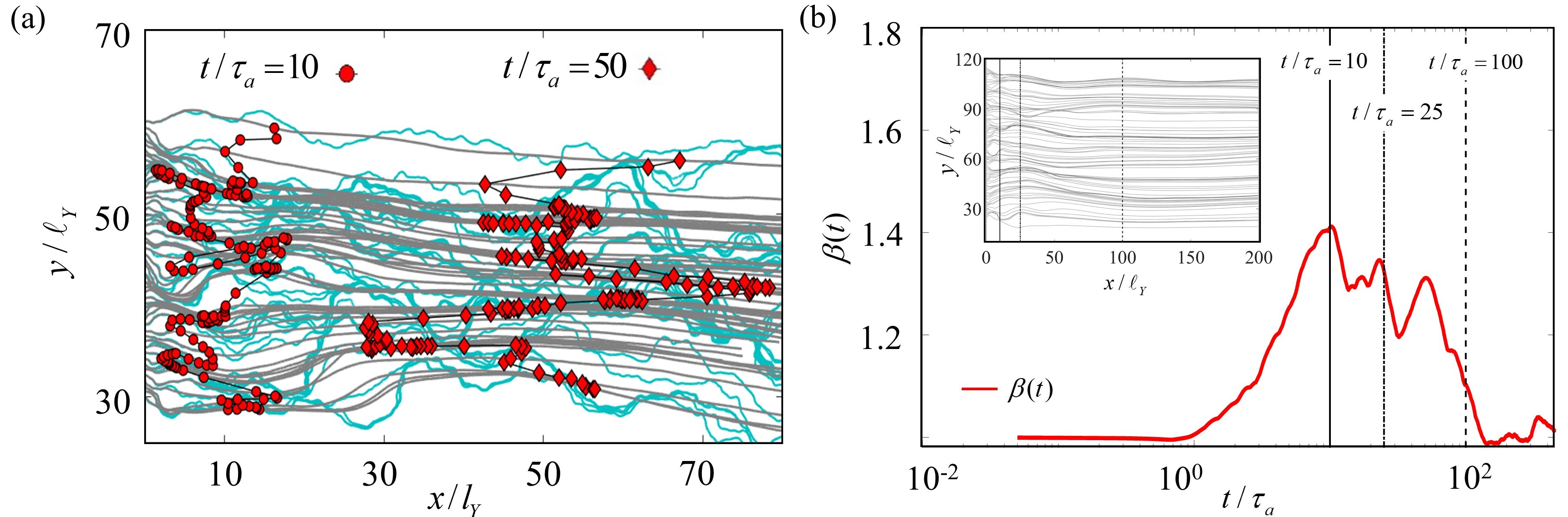}
\end{center}
\caption{(a) Green functions centroids trajectories $\vx^g(t|y')$ (grey curves) and the advective streamlines (light blue curves) counterpart. Snapshots of $\vx^g(t|y')$ are depicted at $t/\tau_a=(10, 50)$ (red symbols).  The distance $\beta(t)$ versus $t/\tau_a$ and $\vx^g(t|y')$ (grey curves) over time (inset). 
} 
\label{fig:Coalescence}
\end{figure}

The presence of high heterogeneity in the formation enhances the focusing of flow into the preferential paths. Thus, we expect even the distance between GFs to be markedly affected by the flow organization in highly heterogeneous formations. Figure \ref{fig:Coalescence}(a) depicts a set of GFs centroids $\vx_{1}^g(t|y')$ at times $t/\tau_a=(10, 50)$ (red symbols) jointly with the whole $\vx_{1}^g(t|y')$ trajectories (grey curves) and the corresponding advective streamlines (light blue curves) originating at the same $y'$ initial positions of the depicted GFs. Inspection of \ref{fig:Coalescence}(a) highlights the close similarity between GFs centroids trajectories and the advective streamlines over the initial traveled distances, while the former becomes straighter as each GF samples the flow variability. Notably, the initial advection-dominated convey of nearby GFs into the closest preferential flow paths sets (\textit{a}) a significant transverse overlapping of some initially close GFs within a sub-set and (\textit{b}) an increase in the transverse distance between GFs pertaining to different subsets. These features clearly affect the mixing dynamics of the plume. Considering the whole set of GFs-pairs initially separated by $\Delta_y$, we evaluate their average transverse distance over time as 
\begin{align}
\label{eq:beta}
\beta(t)=\frac{\overline{ y_{1}^g(t|y'+\Delta_y)-y_{1}^g(t|y')}}{\Delta_y} 
\end{align} 
Figure \ref{fig:Coalescence} (b) depicts $\beta(t)$, given $\Delta_y=\ell_Y$, versus time and in the inset a set of GFs trajectories. Inspection Figure \ref{fig:Coalescence} (b) reveals the initial growth of $\beta(t)$ until $t/\tau_a=(10)$ consistently with the previously discussed behavior of $CV_{\sigma_x^g}(t)$ under the influence of the advection-dominated convey of compact GFs into preferential flow paths. Afterward, $\beta(t)$ decreases exhibiting some peaks that, overall, can be related to the spatial meandering behavior of the trajectories $y_1^g(t|y')$. At the late time, $\beta(t)\approx 1$ due to the sampling of the flow fluctuations which straighten the GFs centroids trajectories over time. Note that, $\beta(t)$ does not quantify the degree of overlap between two GFs initially separated by $\Delta_y=\ell_Y$, but it measures the (average) area swapped by a GF in relation to its two neighboring GFs. Thus, $\beta(t)$ imbues information about the dynamics of coalescence between a GF and its neighbors. The latter aspect is not encompassed in the dispersive lamellae in which GFs do not interact as their relative transverse distances remain constant over time.

\section{Conclusions}
\label{section: Conclusions}
We study the upscaling of the mixing state of a large plume traveling through Darcy's scale heterogeneous formation, by exploring the connection between the dispersive and mixing dynamics of a plume. In particular, we test the dispersive lamellae approach proposed by \citealt{Perez2019} for randomly heterogeneous Darcy's scale formations. The latter approximates the transport Green functions (GFs) as a Gaussian concentration profile that dilutes according to the effective dispersive scale $\sigma^e_x(t)$. This conceptualization leads to satisfactory prediction of the mixing state in case of mild degrees of heterogeneity. On the other hand, the dispersive lamella approach fails to properly predict mixing in highly heterogeneous formations. We attribute this shortcoming to the incapability of the dispersive lamella to account for (\textit{i}) the marked variability within the plume of the GFs dispersive scales that deviate from $\sigma^e_x(t)$ and (\textit{ii}) the convey of initially close GFs into focused preferential flow paths that control the GFs interactions, over the pre-asymptotic times. These findings are guiding the ongoing work to extend the dispersive lamellae for mixing in highly heterogeneous media introducing the sub-plume variability of the GFs dispersive behaviour and a physics-based coalescence model framed in term of the dynamics of the GFs centroids.

\section{Appendix A: Dynamic Binning}
\label{section:CBinning}
The transport problem in Equation \ref{eq:DarcyTrans} is formulated in a Lagrangian framework according to the Langevin equations
\begin{align}
\label{eq:Langevin}
\frac{d\vx_p(t)}{dt}=\textbf{\textit{v}}(\vx_p(t)) + \sqrt{2D}
\mathbf{\xi}(t)
\end{align}
where $\vx_p$ are the solute particles coordinates. Equation \ref{eq:Langevin} is discretized as 
\begin{align}
\label{eq:LangevinDISCRETE}
\vx_p(t+\Delta t)=\vx_p(t)+\textbf{\textit{v}}(\vx_p(t))\Delta t + \sqrt{2D\Delta t} \mathbf{\zeta}(t)
\end{align}
where the noise $\mathbf{\zeta}(t)$ is an uncorrelated standard Gaussian variable. 
In order to obtain the concentration field $c(\vx,t)$ we employ $N_p=10^6$ solute particles and a variable time step of $\Delta t /\tau_a= 5 \times 10^{-3}$ until $t/\tau_a=20$ and $\Delta t/\tau_a = 5 \times 10^{-2}$ afterward. The solute concentration field is then evaluated by employing an adaptive binning procedure. At a given time, we define the square bin edge as $\Delta l_C (t) = min(\sigma_x(t)/10; \sigma_y(t)/10)$ where 
\begin{align}
\label{eq:sigmay}
\sigma_y(t)= \sqrt{ \frac{1}{L_y} \int\limits_{L_y}dy' \left( y_p(t,y')-y_p(t=0,y') \right)^2 }
\end{align}
is the transverse dispersive scale for the plume particles trajectories when the latter are initialized at the same origin. Note that, in Equation \ref{eq:sigmay} we exploit the fact that the average transverse position is null, i.e., $1/L_y \int\limits_{L_y}dy' (y_p(t,y')-y_p(t=0,y'))=0$. We consider a regular binning grid that covers the whole extension of the plume in the longitudinal and transverse directions. The solute concentration field $C(\vx,t)$ is then obtained by evaluating the density of the particles within the dynamic spatial grid as
\begin{align}
\label{eq:CBins}
c(\vx, t) = \frac{1}{N_p} \sum_{n=1}^{N_p} \prod_{i=1}^{2} \delta_{\lfloor \frac{x_i}{\Delta l_C (t)} \rfloor ,{\lfloor \frac{x_i^{n}(t)}{\Delta l_C (t)} \rfloor}} \Delta l_C (t)
\end{align}
where $\lfloor x \rfloor$ is the floor function defined as $\lfloor x \rfloor = max(n \in \mathbb{Z}| n \leq x)$. This simple strategy allows to reprodue the internal solute plume organization satisfactorily while containing the computational costs. Note that, adaptive mesh strategies can be adopted to evaluate the mixing state of a plume employing an Eulerian numerical method (e.g., \citealt{DellOca2018})).

\section{Appendix B: Dynamic Binning}
\label{section:MomentEquation}
We consider the evolution equation for a general function $\phi(\vx,t) = F[c(\vx,t)]$ of the solute concentration. The time-derivative of $\phi(\vx,t)$ can be written as
\begin{align}
    \frac{\partial \phi(\vx,t)}{\partial t} = \frac{d F(x)}{dc} \frac{\partial c(\vx,t)}{\partial t}.  
\end{align}
By using the advection-dispersion equation~\eqref{eq:DarcyTrans}, we can write
\begin{align}
    \frac{\partial \phi(\vx,t)}{\partial t} = \frac{d F(c)}{dc} \left[- \vu(\vx) \cdot \nabla c(\vx,t) + D \nabla^2 c(\vx,t)\right]
\end{align}
The latter can be written as
\begin{align}
    \frac{\partial \phi(\vx,t)}{\partial t} + \vu(\vx) \cdot \nabla \phi(\vx,t) - D \nabla^2 \phi(\vx,t) = - D [\nabla c(\vx,t)]^2 \frac{d^2 F(c)}{dc^2} 
\end{align}
That is, for $\phi(\vx,t) = c(\vx,t)^2$, we obtain the evolution equation
\begin{align}
\label{eq:evolc2}
    \frac{\partial c(\vx,t)^2}{\partial t} + \vu(\vx) \cdot \nabla \phi c(\vx,t)^2 - D \nabla^2 c(\vx,t)^2 = - 2 D [\nabla c(\vx,t)]^2. 
\end{align}
Integration of Equation \ref{eq:evolc2} over space gives 
\begin{align}
\label{eq:scalAppB}
    \frac{d M(t)}{dt} = - 2 \chi(t);   && \chi(t) =\int d\vx  D[\nabla c(\vx,t)]^2
\end{align}

\bibliography{ref}

\end{document}